\documentclass[aps,prl,twocolumn,superscriptaddress]{revtex4}

\usepackage[latin9]{inputenc}
\usepackage{amsmath}
\usepackage{amssymb}
\usepackage{graphicx}
\usepackage{makecell}
\usepackage{array}
\usepackage{CJK}

\usepackage[unicode=true,pdfusetitle,
 bookmarks=false,
 breaklinks=false,pdfborder={0 0 1},backref=false,colorlinks=false]
 {hyperref}
\hypersetup{
 bookmarksnumbered=false,bookmarksopen=false}

\makeatletter
\@ifundefined{textcolor}{}
{%
 \definecolor{BLACK}{gray}{0}
 \definecolor{WHITE}{gray}{1}
 \definecolor{RED}{rgb}{1,0,0}
 \definecolor{GREEN}{rgb}{0,1,0}
 \definecolor{BLUE}{rgb}{0,0,1}
 \definecolor{CYAN}{cmyk}{1,0,0,0}
 \definecolor{MAGENTA}{cmyk}{0,1,0,0}
 \definecolor{YELLOW}{cmyk}{0,0,1,0}
}

\usepackage{color}%\usepackage{amstext}

\newcommand{\ehbar}{\hbar_{\mathrm{eff}}}

\@ifundefined{textcolor}{}{%
 \definecolor{BLACK}{gray}{0}
 \definecolor{WHITE}{gray}{1}
 \definecolor{RED}{rgb}{1,0,0}
 \definecolor{GREEN}{rgb}{0,1,0}
 \definecolor{BLUE}{rgb}{0,0,1}
 \definecolor{CYAN}{cmyk}{1,0,0,0}
 \definecolor{MAGENTA}{cmyk}{0,1,0,0}
 \definecolor{YELLOW}{cmyk}{0,0,1,0}
}

\usepackage{txfonts}

\makeatother

\begin{document}

\title{Quantum criticality at the boundary of the non-Hermitian regime of a Floquet system}

\author{Wen-Lei Zhao}
\email[]{wlzhao@jxust.edu.cn}
\affiliation{School of Science, Jiangxi University of Science and Technology, Ganzhou 341000, China}

\author{Jie Liu}
\email[]{jliu@gscaep.ac.cn}
\affiliation{Graduate School of China Academy of Engineering Physics, Beijing 100193, China}
\affiliation{CAPT, HEDPS, and IFSA Collaborative Innovation Center of the Ministry of Education, Peking University, Beijing 100871, China}

\begin{abstract}
We investigate both analytically and numerically the dynamics of quantum scrambling, characterized by the out-of-time ordered correlators (OTOCs), in a non-Hermitian quantum kicked rotor subject to quantum resonance conditions. Analytical expressions for OTOCs as a function of time are obtained, demonstrating a sudden transition from the linear growth to quadratic growth when the non-Hermitian parameter decays to zero. At this critical point, the rates of the linear growth are found to diverge to infinity, indicating the existence of quantum criticality at the boundary of the non-Hermitian regime.
The underlying mechanism of this quantum criticality is uncovered, and possible applications in quantum metrology are discussed.
\end{abstract}
\date{\today}

\maketitle

{\color{blue}\textit{Introduction.---}}
Uncovering new phases and determining the scaling laws of phase transition
at the critical points are the primary goals of physics~\cite{Cardy96}. Quantum criticality, characterized by nonanalytical behavior of observables, arises from the singularity of energy band landscape~\cite{Sachdev11}. For instance, Dirac points cause the quantized growth of conductance in electron gases as the magnetic field varies~\cite{QNiu84}. In Floquet systems, critical behavior is determined by the singularity of quasienergy bands~\cite{XLiu22,Naji22prb,Naji22pra,Jafari19,KYang19prb}. It is found that the saddle points in the quasienergy landscape lead to a logarithmic divergence in the density of states, akin to excited state phase transitions in static systems, resulting in a magnetization cusp~\cite{Bastidas14L,Bastidas14A,IGMata21}.
Quantum criticality roots in the emergences of large quantum fluctuations and long-range correlations~\cite{Cardy96,Jafari17A,LWZhou18pra}, whose spatio-temporal propagation can be well quantified by the OTOCs~\cite{Larkin1969,Roberts2016}. Indeed, both theoretical and experimental investigations demonstrate that the OTOCs can be used as order parameter to detect the equilibrium~\cite{HShen17prb,Lewis-Swan20prl,Dag19prl}, dynamical~\cite{MHeyl18prl,Nie20prl,Zamani22}, and topological phase transitions~\cite{QianBin23}.

Non-Hermiticity is an essential feature of different systems~\cite{Ashida20,Hatano96PPL}, including cold atoms~\cite{Keller97prl,Takasu20}, optics~\cite{Longhi10prl,QLin22prl,Gupta19}, and dissipative systems~\cite{Tzortzakakis21,Minganti20,Kadanoff68}. Non-Hermitian degeneracies in absorbing anisotropic crystals determine the transmission of polarized light~\cite{Pancharatnam,Berry94,Berry04}. Non-Hermitian random matrix theory is employed to describe the spontaneous breaking of chiral symmetry in quantum chromodynamics~\cite{Stephanov96,Markum99}. Both experimental and theoretical investigations have unveiled fundamental concepts, such as the non-Bloch bulk-boundary correspondence in topological phases~\cite{Yao18}, non-Hermitian skin effects~\cite{WTXue22,SLonghi22prb,LHLin20prl}, and nonreciprocal Landau-Zener tunneling~\cite{WYWang22}. The quantum critical phenomenon in non-Hermitian systems is still an elusive issue and gains rising interests recently. Interestingly, the dynamics of OTOCs is governed by the Yang-Lee edge singularity~\cite{LZhai20} and spontaneous $\cal{PT}$-symmetry breaking~\cite{WlZhao23pra,WLZhao23FIP}. Notably, the OTOCs exhibit quantized phenomenon in $\cal{PT}$-symmetric Floquet systems, indicating the emergence of a novel phase~\cite{WlZhao22}.

In this letter, we investigate the quantum critical phenomenon, quantified by the OTOCs, via a non-Hermitian quantum kicked rotor (NQKR) model with complex kicking strength. We analytically obtain the time dependence of various OTOCs under the quantum resonance condition, revealing their linear growth over time for non-Hermitian kicking. In the Hermitian case, the OTOCs may increase quadratically or remain constant, depending on their specific form. Thus, there exists a sudden transition in the dynamics of the OTOCs at the boundary of the non-Hermitian regime. Interestingly, the growth rate of the OTOCs diverges towards infinity as the non-Hermitian parameter decays to zero, providing evidence of the presence of quantum critical behavior. The finding of quantum criticality at the boundary of the non-Hermitian regime has important implications for the foundation of the non-Hermitian extension of quantum physics. Our results suggest that the OTOCs can be used to identify the nonequilibrium dynamical signatures of  quantum phase transition in non-Hermitian Floquet systems, shedding light on the elusive issue of non-Hermitian wave chaos~\cite{Bera22, Shivam23, WLZhao23sym}.

{\color{blue}\textit{Model and main results.---}}
The Hamiltonian of the NQKR model reads
\begin{equation}\label{Hamil}
{\rm H}=\frac{{p}^2}{2}+ V_K(\theta)\sum_n
\delta(t-t_n)\:,
\end{equation}
with the kicking potential
\begin{equation}\label{NHKicking}
V_K(\theta)= (K+{i} \lambda)\cos(\theta)\;,
\end{equation}
where $p=-i\ehbar\partial/\partial \theta$ is the angular momentum operator, $\theta$ is the angle coordinate, satisfying the communication relation $[\theta,p]=i\ehbar$ with $\ehbar$ effective Planck constant. Here, the parameters $K$ and $\lambda$ control the strength of the real and imaginary parts of the kicking potential, respectively~\cite{Satija02pre,LWZhou19pra}. All variables are properly scaled, thus, in dimensionless unites. The eigenequation of angular momentum operator is $p|n\rangle = p_n |n \rangle$ with eigenvalue $p_n = n\ehbar$ and eigenstate $\langle \theta |n\rangle=e^{in\theta}/\sqrt{2\pi}$. With this complete basis, an arbitrary state can be expanded as $|\psi \rangle=\sum_n \psi_n |n\rangle$. The advancements of delta-kicking potential is that the corresponding Floquet operator can be split into two components, namely $U=U_fU_K$, with the free evolution operator $U_f = \exp(-ip^2/2\ehbar)$ and the kicking term $U_K = \exp[-iV_K(\theta)/2\ehbar]$. The time evolution of a quantum state from $t_n$ to $t_{n+1}$ is governed by $|\psi(t_{n+1})\rangle = U|\psi(t_{n})\rangle$.

The OTOCs are defined as $C(t)=-\langle[A(t),B]^2\rangle$, where both $A(t)=U^{\dagger}(t)A U(t)$ and $B$ are operators evaluated in Heisenberg picture, and $\langle \cdot\rangle=\langle \psi(t_0)|\cdot|\psi(t_0)\rangle$ denotes the average over an initial state $|\psi(t_0)\rangle$~\cite{Li2017,Hashimoto2017,Mata2018,Zonnios22,WLZhao21prb}. We employ different operators to construct the OTOCs. In the first case, we consider the combination of a unitary operator $A=e^{i\varepsilon p}$ and a projection operator $B=|\psi(t_0)\rangle \langle \psi(t_0)|$, which results in the OTOCs $C_{f}=1- |\langle \psi(t)|e^{i\varepsilon p}|\psi(t)\rangle|^2$. In the second case, we utilize the operators $A=p$ and $B=\theta$, yielding $C_{p}=-\langle [p(t),\theta]^2\rangle$. Our main findings can be summarized by the following relationships
\begin{align}\label{FOTCLGwh}
C_{f}\approx
\begin{cases}
\frac{2\pi\varepsilon^2\left(K^2+\lambda^2\right)}{\lambda}t\;, & \text{for $\lambda>0$, $t\gg 1/\lambda$}\;,\\
\frac{\varepsilon^2 K^2 t^2}{2}\;,  & \text{for $\lambda=0$}\;,
\end{cases}
\end{align}
and
\begin{align}\label{PXOTCLGwh}
C_{p}\approx
\begin{cases}
\frac{2\pi^3\left(K^2+\lambda^2\right)}{\lambda}t\;, & \text{for $\lambda>0$, $t\gg 1/\lambda$}\;,\\
16\pi^2\;,  & \text{for $\lambda=0$}\;.
\end{cases}
\end{align}
These relations clearly demonstrate the non-Hermiticity-induced linear increase of both $C_{f}(t)$ and $C_{p}(t)$ with time. Notably, the growth rate $G = dC_{f}/dt$ (or $dC_{p}/dt$) takes the form
\begin{align}\label{GRate}
G \approx
\begin{cases}
\frac{2\pi\varepsilon^2\left(K^2+\lambda^2\right)}{\lambda}\;, & \text{for $C_f$}\;,\\
\frac{2\pi^3\left(K^2+\lambda^2\right)}{\lambda}\;, & \text{for $C_p$}\;,
\end{cases}
\end{align}
which displays the remarkable divergence of the $G$ as $\lambda$ approaches zero.

\begin{figure}[t]
\begin{center}
\includegraphics[width=8cm]{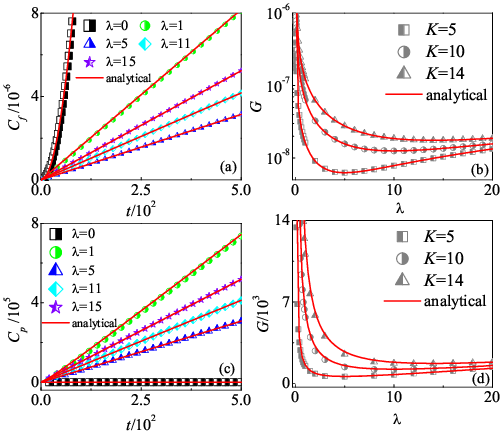}
\caption{Left panels: Time dependence of the $C_f$ (a) and $C_p$ (c) with $K=5$ for $\lambda=0$ (squares), $1$ (circles), $5$ (triangles), $11$ (diamonds), and $15$ (pentagram). In (a) and (c): Red lines indicate the theoretical prediction in Eqs.~\eqref{FOTCLGwh} and ~\eqref{PXOTCLGwh}, respectively. Right panels: The growth rate $G$ of the $C_f$ (b) and $C_p$ (d) versus $\lambda$ for $K=5$ (squares), $10$ (circles), and $14$ (triangles). In (b) and (d): Red lines indicate the theoretical prediction in Eq.~\eqref{GRate}. The parameters are $\ehbar=4\pi$ and $\varepsilon=10^{-5}$.\label{OTCCritical}}
\end{center}
\end{figure}

To verify the theoretical predictions mentioned above, we numerically investigate the time dependence for both $C_f$ and $C_p$ over a wide range of $\lambda$. It is worth noting that our system does not possess well-defined thermal states, as the temperature of periodically-driven systems tends to increase indefinitely over time~\cite{D'Alessio14}. Consequently, there is no need to perform thermal averaging when calculating the OTOCs. Without loss of generality, we select the ground state as the initial state in numerical simulations, i.e., $|\psi(t_0)\rangle=1/\sqrt{2\pi}$. Figure~\ref{OTCCritical}(a) shows that the $C_f$ increases in the quadratic function of time for Hermitian case, i.e., $\lambda=0$. Interestingly, for sufficiently large $\lambda$ (e.g., $\lambda=1$), $C_f$ exhibits a fascinating linear growth. Both the linear and quadratic behaviors are in perfect consistency with our theoretical predictions in Eq.~\eqref{FOTCLGwh}. We further investigate the growth rate $G$ of $C_f$ for different $\lambda$. Figure~\ref{OTCCritical}(b) illustrates that, for a specific $K$ (e.g., $K=5$), the $G$ initially decreases monotonically to a minimum value with the increase of $\lambda$, and then it increases. This behavior of $G$ is in good agreement with our theoretical prediction in Eq.~\eqref{GRate}, which suggests a divergence of $G$ as $\lambda$ approaches zero. For $\lambda=0$, the $C_p$ remains constant over time, while for sufficiently large $\lambda$ (e.g., $\lambda=1$ in Fig.\ref{OTCCritical}(c)), it linearly increases with time, in accordance with the laws described in Eq.\eqref{PXOTCLGwh}. The corresponding growth $G$ demonstrates the divergence of $G$ with $\lambda \rightarrow 0$, confirming the validity of the theoretical prediction in Eq.~\eqref{GRate} [see Fig.~\ref{OTCCritical}(d)].

The threshold value  of the non-Hermitian parameter, determined by the appearance of complex quasienergies, is usually not identical to the mathematical boundary of the non-Hermitian regime~\cite{CTWest10prl, WLZhao23sym, WlZhao23pra, KQHuang21}. While recently, in a $\mathcal{PT}$-symmetric kicked rotor model under the quantum resonance
condition, it was found numerically that the complex quasienergies emerge exactly at the boundary of the non-Hermitian regime~\cite{Longhi2017}. Our analytical deduction explicitly demonstrates that this holds true for a general non-Hermitian Floquet system under the quantum resonance condition, and the underlying quantum critical phenomenon can be well depicted by the behaviors of the time evolution of the OTOCs at the non-Hermitian boundary.

%%%%%%%%%%%%%%%%%%%%%%%%

{\color{blue}\textit{Theoretical analysis.---}}
As an illustration, we provide detailed derivations of $C_f$ below~\cite{SupplMaterial}. It is straightforward to obtain the relation
\begin{equation}\label{otoc}
C_f(t)= C_1(t)+ C_2(t)-2{\rm Re}\left[C_3(t)\right]\;,
\end{equation}
where the two-point correlators are defined as
\begin{align}\label{FPart}
C_{1}(t)& \mathrel{\mathop:}= \langle A^{\dagger}(t)B^2A(t) \rangle=\langle \psi_R(t_0) | B^2 |\psi_R(t_0)\rangle\;,
\end{align}
\begin{align}\label{SPart}
C_{2}(t) &\mathrel{\mathop:}= \langle B^{\dagger} A^{\dagger}(t)A(t)B\rangle =\langle \varphi_R(t_0)|\varphi_R(t_0)\rangle\;,
\end{align}
and the four-point correlator
\begin{align}\label{TPart}
C_{3}(t) &\mathrel{\mathop:}= \langle A^{\dagger}(t)B A(t)B \rangle=\langle \psi_R(t_0)|B|\varphi_R(t_0)\rangle\;.
\end{align}
Here, $|\psi_R(t_0)\rangle = U^{\dagger}(t)AU(t)|\psi(t_0)\rangle$ and $|\varphi_R(t_0)\rangle =U^{\dagger}(t)AU(t)B|\psi(t_0)\rangle$ represent the states at the end of time reversal.

Given $A = e^{i\varepsilon p}$ and $B = |\psi(t_0)\rangle\langle \psi(t_0)|$, we obtain the equivalences $C_{1}(t) = C_{3}(t) = |\langle \psi(t) | e^{i\varepsilon p} |\psi(t)\rangle|^2$ and $C_2(t) = \langle \psi(t)|e^{-i\varepsilon p}U(t)U^{\dagger}(t)e^{i\varepsilon p}|\psi(t)\rangle = |\langle \psi(t)|\psi(t)\rangle|^2$. Consequently, $C_f(t) = |\langle \psi(t)|\psi(t)\rangle|^2 - |\langle \psi(t) | e^{i\varepsilon p} |\psi(t)\rangle|^2$, with the latter term referred to as fidelity out-of-time-ordered correlators (FOTOC), i.e., $\mathcal{F}_o = |\langle \psi(t) | e^{i\varepsilon p} |\psi(t)\rangle|^2$~\cite{Yan20prl,LewisSwan19,WLZhao23arx}. It should be noted that the norm $\mathcal{N}(t) = \langle \psi(t)|\psi(t)\rangle$ of non-Hermitian systems unboundedly increases for sufficiently large non-Hermitian parameters. To remove the influence of the norm, we define the rescaled FOTOC as $\mathcal{F}_o(t) = |\langle \psi(t) | e^{i\varepsilon p} |\psi(t)\rangle/\mathcal{N}(t)|^2$, thus yielding $C_f(t) = 1 - \mathcal{F}_o(t)$. Considering the approximation $e^{i\varepsilon p}\approx 1+ i\varepsilon p$ for $\varepsilon \ll 1$, the FOTOC can be described as $\mathcal{F}_o(t)\approx 1-\varepsilon^2 \left[\langle\psi(t)| p^2|\psi(t)\rangle -\langle\psi(t)| p|\psi(t)\rangle^2\right]$. Therefore, we have
\begin{equation}\label{Fotoc0}
C_f(t)\approx \varepsilon^2 \left[\langle\psi(t)| p^2|\psi(t)\rangle -\langle\psi(t)| p|\psi(t)\rangle^2\right]\;.
\end{equation}

In the main quantum resonance condition $\ehbar=4\pi$, each matrix element of $U_f$ equals unity, i.e., $U_f(n)=\exp(-i n^2\hbar/2)=1$, rendering it ineffective in the time evolution of quantum states. The quantum state at any given time $t$ can be obtained by multiplying the initial state with the kicking evolution operator $t$ times, i.e., $|\psi(t)\rangle = U_K^{t}|\psi(t_0)\rangle$. By setting the ground state as the initial state [i.e., $\psi(t_0)=1/\sqrt{2\pi}$], the quantum state $|\psi(t)\rangle$ in coordinate space takes the form
\begin{equation}\label{QStateTn}
\psi(\theta,t)=\frac{1}{\sqrt{2\pi}}\exp{\left\{-\frac{it}{4\pi}\left[(K+{i} \lambda)\cos(\theta)\right]\right\}}\:,
\end{equation}
where the norm $\mathcal{N}(t)=I_0\left({\lambda t}/{2\pi}\right)$ and $I_m(x)$ denotes the modified Bessel function of order $m$~\cite{SupplMaterial}.
It is worth noting that, for $\lambda t/2\pi \gg 1$, the norm $\mathcal{N}(t)\approx\exp(\lambda t/2\pi)/\sqrt{\lambda t}$ increases unboundedly with time even for arbitrarily small $\lambda$, indicating that the quasienergies are complex in quantum resonance case~\cite{Longhi2017}.

The even symmetry of the state in Eq.\eqref{QStateTn} results in zero mean momentum, i.e., $\langle p(t)\rangle =0$. Lengthy but straightforward derivations yield the rescaled mean square of momentum by dividing the norm
\begin{equation}\label{MEnergy}
\langle p^2(t)\rangle = 2\pi t\frac{I_1\left(\frac{\lambda t}{2\pi}\right)}{I_0\left(\frac{\lambda t}{2\pi}\right)}\frac{K^2+\lambda^2}{\lambda}\;.
\end{equation}
Combining Eqs.\eqref{MEnergy} and ~\eqref{Fotoc0}, we obtain the relation
\begin{equation}\label{Fotoc1}
C_f(t)\approx 2\pi \varepsilon^2t\frac{I_1\left(\frac{\lambda t}{2\pi}\right)}{I_0\left(\frac{\lambda t}{2\pi}\right)}\frac{K^2+\lambda^2}{\lambda}\;.
\end{equation}

Consider two limits: ${\lambda t}/{2\pi}\ll 1$ and ${\lambda t}/{2\pi}\gg 1$. In the ${\lambda t}/{2\pi}\ll 1$ limit, where $\lambda\ll 1$ and $t\ll 1/\lambda$, we can make the estimations
\begin{equation}\label{SmallApp}
I_1\left(\frac{\lambda t}{2\pi}\right)\approx
\frac{1}{\Gamma(2)}\frac{\lambda t}{4\pi}=\frac{\lambda t}{4\pi}\;,\; \text{and}\;\;  I_0\left(\frac{\lambda t}{2\pi}\right)\approx
\frac{1}{\Gamma(1)}=1\;,
\end{equation}
where $\Gamma (x)$ is the Gamma function with $\Gamma(2)=1$ and $\Gamma(1)=1$. Thus, $C_f$ can be approximated as
\begin{equation}\label{Fotoc2}
C_f(t) \approx \frac{\varepsilon^2K^2t^2}{2}\;, \quad \text{with}\quad \lambda=0\;.
\end{equation}
For $\frac{\lambda t}{2\pi}\gg 1$, the relation $I_0\left(\frac{\lambda t}{2\pi}\right) \approx I_1\left(\frac{\lambda t}{2\pi}\right)$ yields
\begin{equation}\label{Fotoc3}
C_f(t) \approx 2\pi \varepsilon^2 t \frac{K^2+\lambda^2}{\lambda}\;.
\end{equation}
Both the Eq.~\eqref{Fotoc2} and Eq.~\eqref{Fotoc3} have been summarized in Eq.~\eqref{FOTCLGwh}.

{\color{blue}\textit{Conclusion and discussions.---}}
In this work, we investigate the dynamics of various OTOCs, including the FOTOC, via a NQKR model with quantum resonance condition. Analytical functions of the OTOCs on both the system parameters and time are achieved and confirmed by our numerical results as well. It demonstrates the linear growth of various OTOCs with time in the non-Hermitian regime, with a remarkable divergence of the growth rate at the zero value of the imaginary part of the kicking potential. This unveils the existence of a Floquet-driving induced quantum criticality at the boundary between Hermitian and non-Hermitian regimes. In recent years, the rich physics exhibited by Floquet systems has garnered increasing attention,  which spurs potential applications of Floquet engineering such as manipulating the Anderson metal-insulator transition~\cite{JChab08prl,Tian11prl}, many-body dynamical localization~\cite{Nag14,Tamang21,Chiew23}, and topological phases~\cite{LWZhou18prb,LWZhou23arx,Dahlhaus11,Nieuwenburg12,Dana19,ZYCheng22,KYang22,WWZhu21,LHLi18,Ho12,LWZhou21prr}. In this context, our findings represent a significant advancement in exploring novel phases within non-Hermitian Floquet systems~\cite{HYWang22}. Given that the FOTOC is directly proportional to the quantum Fisher Information~\cite{Garttner18prl}, the divergence of the FOTOC at $\lambda=0$ holds practical significance in enhancing the measurement precision of quantum sensing~\cite{SCLi21,Chu20,Wiersig16,Liu16}.

Based on the mathematical equivalence between the propagation equation of light under paraxial approximations and the Schr{\" o}dinger equation~\cite{Longhi09}, quantum simulation using optical setups has become an active field for experimentally confirming fundamental concepts of quantum mechanics, including topologically protected transport~\cite{Ozawa19,LXiao20,Smirnova20,YShen22}, Anderson localization~\cite{Wiersma97,Segev13,Schirmacher18}, and Bloch oscillations~\cite{Longhi08prl,Sapienza03prl}. The paradigmatic kicked rotor model has been realized using all-optical systems, where the periodically modulated refractive index
of optical fibers~\cite{Prange89,Agam92} or the periodically arranged phase gratings with equal distance~\cite{Fischer00,Rosen00} mimic the delta-kicking potential for a quantum particle. The well-known dynamical localization of light in the frequency domain has been observed. Notably, the quantum resonance condition can be achieved by tuning the distance between phase gratings, allowing for experimental realizations of both ballistic diffusion and ratchet transport in momentum space~\cite{CZhang15}. It is worth noting that the loss feature of phase gratings has been utilized to emulate the $\cal{PT}$-symmetric kicked rotor model~\cite{Longhi2017}. Therefore, the NQKR model is achievable in state-of-the-art optical experiments, in which the FOTOC, proportional to the mean energy, can be measured in frequency domain.

{\color{blue}\textit{Acknowledgments.---}} Wen-Lei Zhao is supported by the National Natural Science Foundation of China (Grant Nos. 12065009), the Natural Science Foundation of Jiangxi province (Grant Nos. 20224ACB201006 and 20224BAB201023) and the Science and Technology Planning Project of Ganzhou City (Grant No. 202101095077). Jie Liu is supported by the NSAF (Contract No. U1930403).

\pagebreak
\clearpage
%\newpage
\widetext

\begin{center}
\textbf{\large Supplemental Material:\\ Quantum criticality at the boundary of the non-Hermitian regime of a Floquet system}
\end{center}
%%%%%%%%%% Merge with supplemental materials %%%%%%%%%%
%%%%%%%%%% Prefix a "S" to all equations, figures, tables and reset the counter %%%%%%%%%%
%%%%%%%%%% Merge with supplemental materials %%%%%%%%%%
%%%%%%%%%% Prefix a "S" to all equations, figures, tables and reset the counter %%%%%%%%%%
\setcounter{equation}{0} \setcounter{figure}{0} \setcounter{table}{0}
\setcounter{page}{1} \makeatletter \global\long\def\theequation{S\arabic{equation}}
 \global\long\def\thefigure{S\arabic{figure}}
 \global\long\def\bibnumfmt#1{[S#1]}
 \global\long\def\citenumfont#1{S#1}
%\tableofcontents{}

%\begin{center}
In this supplementary material, we present a numerical method for computing out-of-time-ordered correlators (OTOCs) in non-Hermitian systems. Additionally, we provide detailed derivations of $C_p=-\langle[p(t),\theta]^2 \rangle$ in the non-Hermitian quantum kicked rotor (NQKR) model, considering the quantum resonance condition $\ehbar=4\pi$.
%\end{center}

\section*{S.1 Numerical method for calculating OTOCs}
The OTOCs are defined as $C(t)=-\langle|[A(t),B]|^2\rangle$ with $A(t)=U^{\dagger}(t)AU(t)$~\cite{Li2017SM,Hashimoto2017SM,Mata2018SM,Zonnios22SM}.
The decomposition of OTOCs can be expressed as
\begin{equation}\label{otoc}
C(t)= C_1(t)+ C_2(t)-2{\rm Re}\left[C_3(t)\right]\;,
\end{equation}
with
\begin{align}\label{FPart}
C_{1}(t)& \mathrel{\mathop:}= \langle A^{\dagger}(t)B^2A(t) \rangle=\langle \psi_R(t_0) | B^2 |\psi_R(t_0)\rangle\;,
\end{align}
\begin{align}\label{SPart}
C_{2}(t) &\mathrel{\mathop:}= \langle B^{\dagger} A^{\dagger}(t)A(t)B\rangle =\langle \varphi_R(t_0)|\varphi_R(t_0)\rangle\;,
\end{align}
and
\begin{align}\label{TPart}
C_{3}(t) &\mathrel{\mathop:}= \langle A^{\dagger}(t)B A(t)B \rangle=\langle \psi_R(t_0)|B|\varphi_R(t_0)\rangle\;,
\end{align}
where $|\psi_R(t_0)\rangle = U^{\dagger}(t)AU(t)|\psi(t_0)\rangle$ and $|\varphi_R(t_0)\rangle =U^{\dagger}(t)AU(t)B|\psi(t_0)\rangle$ represent the states at the end of time reversal~\cite{WLZhao21prbSM}.

There are four steps for calculating $C_1$ for a specific time $t=t_n$:

1). At the initial time $t=t_0$, we choose an initial state $|\psi(t_0)\rangle$ with a unity norm $\mathcal{N}_{\psi}(t_0)= \langle \psi(t_0)|\psi(t_0)\rangle=1$.

2). From $t=t_0$ to $t=t_n$, forward time evolution yields the state
\begin{equation}\label{Evolution-c1}
|\psi(t_n)\rangle =U(t_n)|\psi(t_0)\rangle,
\end{equation}
with the norm $\mathcal{N}_{\psi}(t_n)= \langle \psi(t_n)|\psi(t_n)\rangle$. Non-unitary evolution of non-Hermitian systems leads to the growth of the norm, i.e., $\mathcal{N}_{\psi}(t_n)>\mathcal{N}_{\psi}(t_0)$. Accordingly, we use the ratio $\mathcal{F}_{\psi}=\mathcal{N}_{\psi}(t_n)/\mathcal{N}_{\psi}(t_0)$ to measure the increase in the norm.

3). At time $t=t_n$, we apply the operator $A$ to the state $|\psi(t_n)\rangle$
\begin{equation}\label{Aact-c1}
|\tilde{\psi}(t_n)\rangle =A|\psi(t_n)\rangle,
\end{equation}
with the norm $\tilde{\mathcal{N}}_{\psi}(t_n)= \langle \tilde{\psi}(t_n)|\tilde{\psi}(t_n)\rangle=\langle \psi(t_n)|A^2|\psi(t_n)\rangle$, which is the expectation value of the operator $A^2$.

4). From $t=t_n$ to $t=t_0$, backward evolution (i.e., time reversal) starting from the state $|\tilde{\psi}(t_n)\rangle$ yields
\begin{equation}\label{reEvolution-c1}
|\psi_R(t_0)\rangle =U^{\dagger}(t_n)|\tilde{\psi}(t_n)\rangle,
\end{equation}
with the norm $\mathcal{N}_{\psi_R}(t_0)= \langle \psi_R(t_0)|\psi_R(t_0)\rangle$. Similar to step 2), we define the factor $\mathcal{F}_{\psi_R}=\mathcal{N}_{\psi_R}(t_0)/\tilde{\mathcal{N}}_{\psi}(t_n)$ to quantify the growth of the norm.

Since the contribution of the norm to OTOCs is physically meaningless, we define the scaled $C_1(t_n)$ as
\begin{equation}\label{c1-last}
C_1(t_n)=\frac{C_1(t_n)}{\mathcal{F}_{\psi}\mathcal{F}_{\psi_R}} = \frac{\langle\psi_R(t_0)|B^2|\psi_R(t_0)\rangle \tilde{\mathcal{N}}_{\psi}(t_n)}{\mathcal{N}_{\psi}(t_n)\mathcal{N}_{\psi_R}(t_0)}\;.
\end{equation}
This scaling eliminates the growth of the norm during both the forward and backward evolution. In the above equation, we used the relations $\mathcal{N}_{\psi}(t_0)=1$.

Four steps are involved in calculating $C_2$ at a specific time $t = t_n$:

1). At $t = t_0$, the initial state $|\varphi(t_0)\rangle = B|\psi(t_0)\rangle$ is obtained, with a norm $\mathcal{N}_{\varphi}(t_0) = \langle \psi(t_0)|B^2|\psi(t_0)\rangle$, which is the expectation value of $B^2$ over $|\psi(t_0)\rangle$.

2). From $t_0$ to $t_n$, forward time evolution yields $|\varphi(t_n)\rangle = U(t_n)|\varphi(t_0)\rangle$ with a norm $N_{\varphi}(t_n) = \langle \varphi(t_n)|\varphi(t_n)\rangle$. Similar to the step 2 of $C_1(t_n)$, we define a factor $\mathcal{F}_{\varphi} = \mathcal{N}{\varphi}(t_n)/\mathcal{N}_{\varphi}(t_0)$ to measure the norm increase.

3). At $t=t_n$, we apply the operator $A$ to $|\varphi(t_n)\rangle$,
\begin{equation}\label{Aact-c2}
|\tilde{\varphi}(t_n)\rangle =A|\varphi(t_n)\rangle\;,
\end{equation}
with the norm $\tilde{\mathcal{N}}_{\varphi}(t_n)= \langle \tilde{\varphi}(t_n)|\tilde{\varphi}(t_n)\rangle=\langle \varphi(t_n)|A^2|\varphi(t_n)\rangle$.

4). For $t_n\to t_0$, the backward evolution is performed by
\begin{equation}\label{reEvolution-c2}
|\varphi_R(t_0)\rangle =U^{\dagger}(t_n)|\tilde{\varphi}(t_n)\rangle\;,
\end{equation}
with a norm $N_{\varphi_R}(t_0)= \langle \varphi_R(t_0)|\varphi_R(t_0)\rangle$. We use the ratio $\mathcal{F}_{\varphi_R} = \mathcal{N}_{\varphi_R}(t_0)/\tilde{\mathcal{N}}_{\varphi}(t_n)$ to quantify the norm growth during time reversal.

Similar to the scaling of $C_1(t_n)$, we define the rescaled $C_2(t_n)$ as
\begin{equation}\label{numrenormalization-c2t0last}
\begin{aligned}
C_2(t_n)=\frac{C_2(t_n)}{\mathcal{F}_{\varphi}\mathcal{F}_{\varphi_R}}
=\langle\varphi_R(t_0)|\varphi_R(t_0)\rangle\frac{\mathcal{N}_{\varphi}(t_0)\tilde{\mathcal{N}}_{\varphi}(t_n)}{\mathcal{N}_{\varphi}(t_n)\mathcal{N}_{\varphi_R}(t_0)}=\frac{\mathcal{N}_{\varphi}(t_0)}{\mathcal{N}_{\varphi}(t_n)}\tilde{\mathcal{N}}_{\varphi}(t_n)\;.
\end{aligned}
\end{equation}

Given the availability of both $|\psi_R (t_0)\rangle$ and $|\varphi_R(t_0)\rangle$, we can define the rescaled $C_3(t_n)$ as
\begin{equation}\label{TPart-Num}
\begin{aligned}
C_{3}(t_n) = \frac{C_{3}(t_n)}{\sqrt{\mathcal{F}_{\psi}\mathcal{F}_{\psi_R}\mathcal{F}_{\varphi}\mathcal{F}_{\varphi_R}}}=\langle \psi_R(t_0)|B|\varphi_R(t_0)\rangle\sqrt{\frac{\tilde{\mathcal{N}}_{\psi}(t_n)}{\mathcal{N}_{\psi}(t_n)\mathcal{N}_{\psi_R}(t_0)}}
\sqrt{\frac{\mathcal{N}_{\varphi}(t_0)\tilde{\mathcal{N}}_{\varphi}(t_n)}{\mathcal{N}_{\varphi}(t_n)\mathcal{N}_{\varphi_R}(t_0)}}\;,
\end{aligned}
\end{equation}
where we have used the relation $\mathcal{N}_{\psi}(t_0)=1$.

\section*{S.2 The OTOCs of non-Hermitian kicked rotor with resonance}

The Hamiltonian of NQKR reads
\begin{equation}\label{Hamil}
{\rm H}=\frac{{p}^2}{2}+ (K+{i} \lambda)\cos(\theta)\sum_n
\delta(t-t_n)\:,
\end{equation}
where $p=-i \ehbar\partial/\partial \theta$ is the angular momentum operator, $\theta$ is the angle coordinate, $\ehbar$ is the effective Planck constant, $K$ denotes the strength of the real part of the kicking potential, and $\lambda$ indicates the strength of its imaginary part~\cite{Satija02preSM,LWZhou19pra}. All variables are properly scaled, hence in dimensionless units. In quantum resonance condition $\ehbar=4\pi$, the free evolution operator has no effects on quantum states, as its  matrix elements are equal to unity, i.e., $U_f(n)=\exp(-i n^2\ehbar/2)=1$. Therefore, only the kicking evolution, governed by the Floquet operator $U_K(\theta)=\exp\left[-\frac{i}{4\pi}(K+ i\lambda)\cos(\theta)\right]$, needs to be considered. At an arbitrary time $t=t_n$, the quantum state in coordinate space can be precisely expressed as
\begin{equation}\label{QState}
\begin{aligned}
\psi(\theta,t_n)=U_K^{t_n}(\theta)\psi(\theta,t_0)=\exp\left[-\frac{i}{4\pi}(K+i\lambda)t_n\cos(\theta)\right]\psi(\theta,t_0)\:.
\end{aligned}
\end{equation}
The OTOCs, constructed using operators $A=p$ and $B=\theta$, takes the form
\begin{equation}\label{otoc-c}
C_p(t_n)=-\langle[p(t_n),\theta]^2\rangle=C_1(t_n)+C_2(t_n)-2\text{Re}[C_3(t_n)]\;,
\end{equation}
with
\begin{equation}\label{c1SM0}
C_1(t_n)=\langle\psi_R(t_0)|\theta^2|\psi_R(t_0)\rangle\;,
\end{equation}
\begin{equation}\label{c2SM0}
C_2(t_n)=\langle\varphi_R(t_0)|\varphi_R(t_0)\rangle\;,
\end{equation}
and
\begin{equation}\label{c3SM0}
C_3(t_n)=\langle\psi_R(t_0)|\theta|\varphi_R(t_0)\rangle\;.
\end{equation}
Both $|\psi_R(t_0)\rangle =U^{\dagger}(t_n)pU(t_n)|\psi(t_0)\rangle$ and $|\varphi_R(t_0)\rangle =U^{\dagger}(t_n)pU(t_n)\theta|\psi(t_0)\rangle$ represent the states at the end of time reversal.

\subsection*{S.2.1 Detailed derivation of $C_1(t_n)$}

For convenience in analytical derivation, we adopt the uniform ground state $\psi(\theta,t_0)=1/\sqrt{2\pi}$ as the initial state. Based on Eq.~\eqref{QState}, the quantum state can be expressed as
\begin{equation}\label{QuanState-c1tn}
\psi(\theta,t_n)=\frac{1}{\sqrt{2\pi}}\exp\left[-\frac{i}{4\pi}(K+i\lambda)t_n\cos(\theta)\right]\:.
\end{equation}
The norm of this state is given by
\begin{equation}\label{QSnorm-c1tn}
\mathcal{N}_{\psi}(t_n)=\int_{-\pi}^{\pi}d\theta|\psi(\theta,t_n)|^2=I_0\left(\frac{\lambda t_n}{2\pi}\right)\:,
\end{equation}
where $I_0$ represents the modified Bessel function of zeroth order. When $\lambda t_n/2\pi \gg 1$, we can approximate the norm as $\mathcal{N}_{\psi}(t_n)\approx {\exp{(\lambda t_n/2\pi)}}/{\sqrt{\lambda t_n}}$.
At time $t=t_n$, applying the operator $p$ to the state $\psi(\theta,t_n)$ gives
\begin{equation}\label{QuStn-c1}
\tilde{\psi}(\theta,t_n)=p\psi(\theta,t_n)=(K+i\lambda)t_n\sin(\theta)\psi(\theta,t_n)\:,
\end{equation}
with the norm
\begin{equation}\label{QuSnormtn-c1}
	\begin{aligned}
\tilde{N}_{\psi}(t_n)=\langle\psi(t_n)|p^2|\psi(t_n)\rangle
=\frac{2\pi (K^2+\lambda^2)^2 t_n}{\lambda}I_1\left(\frac{\lambda t_n}{2\pi}\right)\:.
\end{aligned}
\end{equation}
Performing time reversal for $t_n\to t_0$ starting from $\tilde{\psi}(\theta,t_n)$ yields
\begin{equation}\label{reQuanState-c1t0}
\begin{aligned}
\psi_R(\theta,t_0)=\left[U^{t_n}_K(\theta)\right]^{\dagger}\tilde{\psi}(\theta,t_n)=\frac{(K+i\lambda)t_n}{\sqrt{2\pi}}\sin(\theta)\exp{\left[\frac{\lambda t_n}{2\pi}\cos(\theta)\right]}\:.
\end{aligned}
\end{equation}
By straightforward calculations, the norm can be obtained as:
\begin{equation}\label{reQuSn2-c1t0}
	\begin{aligned}
\mathcal{N}_{\psi_R}(t_0)=\int_{-\pi}^{\pi}d\theta|\psi_R(\theta,t_0)|^2=\frac{(K^2+\lambda^2)\pi t_n}{\lambda}I_1\left(\frac{\lambda t_n}{\pi}\right)\:.
	\end{aligned}
\end{equation}

Based on Eqs.~\eqref{c1SM0} and ~\eqref{reQuanState-c1t0}, we can derive the expression for $C_1(t_n)$ as follows:
\begin{equation}\label{C1LS-e1}
	\begin{aligned}
C_1(t_n)&=\int_{-\pi}^{\pi}\theta^2|\psi_R(\theta,t_0)|^2d\theta=\frac{|K_{\lambda}|^2t_n^2}{2\pi}
\int_{-\pi}^{\pi}\theta^2\sin^2(\theta)\exp{\left[\frac{\lambda t_n}{\pi}\cos(\theta)\right]}d\theta\:,\\
&\approx 3\pi^2\frac{(K^2+\lambda^2)}{\lambda^2}I_0\left(\frac{\lambda t_n}{\pi}\right)\:,
	\end{aligned}
\end{equation}
where we have used the approximations $\cos(\theta)\approx 1-\theta^2/2$ and $\sin^2(\theta)\approx\theta^2$ for $\lambda t_n/\pi\gg 1$. By substituting Eqs.\eqref{QSnorm-c1tn},\eqref{QuSnormtn-c1},\eqref{reQuSn2-c1t0}, and\eqref{C1LS-e1} into Eq.~\eqref{c1-last}, we obtain the relation
\begin{equation}\label{C1LS-e2}
	\begin{aligned}
C_1(t_n)&=C_1(t_n)\frac{\tilde{\mathcal{N}}_{\psi}(t_n)}{\mathcal{N}_{\psi}(t_n)\mathcal{N}_{\psi_R}(t_0)}
=6\pi^2\frac{(K^2+\lambda^2)}{\lambda^2}\:.
	\end{aligned}
\end{equation}
In Hermitian case $\lambda=0$, it is straightforward to acheive
\begin{equation}\label{C1Hermitian}
	\begin{aligned}
C_1(t_n)=\int_{-\pi}^{\pi}\theta^2|\psi_R(\theta,t_0)|^2d\theta=\frac{K^2t_n^2}{2\pi}
\int_{-\pi}^{\pi}\theta^2\sin^2(\theta)d\theta=\frac{\pi^2 K^2 t_n^2}{6}\:.
	\end{aligned}
\end{equation}

\subsection*{S.2.2 Detailed derivation of $C_2(t_n)$}

The initial state is given by $|\varphi(t_0)\rangle=\theta|\psi(t_0)\rangle={\theta}/{\sqrt{2\pi}}$, with a norm
\begin{equation}\label{INStatenorm-c2}
\mathcal{N}_{\varphi}(t_0)=\int_{-\pi}^{\pi}d\theta|\varphi(\theta,t_0)|^2=\frac{\pi^2}{3}\:.
\end{equation}
The forward evolution from $t_0$ to $t_n$ yields the state
\begin{equation}\label{QTState-c2}
|\varphi(t_n)\rangle=U_K^{t_n}(\theta)|\varphi(t_0)\rangle=U_K^{t_n}(\theta)\theta|\psi(t_0)\rangle=\theta|\psi(t_n)\rangle\:.
\end{equation}
The corresponding norm can be approximated as
\begin{equation}\label{QTStnorm-c2}
\begin{aligned}
\mathcal{N}_{\varphi}(t_n)=\int_{-\pi}^{\pi}d\theta|\varphi(\theta,t_n)|^2=\int_{-\pi}^{\pi}\theta^2|\psi(\theta,t_n)|^2d\theta\approx \frac{2\pi}{\lambda t_n}I_0\left(\frac{\lambda t_n}{2\pi}\right)\:,
\end{aligned}
\end{equation}
where we have used the relation $\cos(\theta)\approx 1-\theta^2/2$ for $2\lambda t_n/\ehbar\gg 1$.
At time $t_n$, applying the operator $p$ to $\varphi(\theta,t_n)$ gives the state
\begin{equation}\label{QuStn-c2}
|\tilde{\varphi}(t_n)\rangle=p|\varphi(t_n)\rangle=\theta|\tilde{\psi}(t_n)\rangle-i4\pi|\psi(t_n)\rangle\:.
\end{equation}
In condition that $\lambda t_n/2\pi \gg 1$, its norm can be described as
\begin{equation}\label{LLLastQuSnormtn-c2}
	\begin{aligned}
	\tilde{N}_{\varphi}(t_n)= \langle \tilde{\varphi}(t_n)|\tilde{\varphi}(t_n)\rangle\approx 12\pi^2\frac{(K^2+\lambda^2)}{\lambda^2}I_0\left(\frac{\lambda t_n}{2\pi}\right)\;.
     \end{aligned}
\end{equation}

For backward evolution from $t_n$ to $t_0$, starting from the state $|\tilde{\varphi}(t_n)\rangle$ in Eq.~\eqref{QuStn-c2}, we obtain
\begin{equation}\label{reQTState-c2-0}
	\begin{aligned}
|\varphi_R(t_0)\rangle=U^{\dagger}(t_n)|\tilde{\varphi}(t_n)\rangle=\theta|\psi_R(t_0)\rangle -i 4\pi|\psi_f(t_0)\rangle\:,
	\end{aligned}
\end{equation}
with $|\psi_f(t_0)\rangle=\left[U_K^{t_n}(\theta)\right]^{\dagger}
|\psi(t_n)\rangle$. Straightforward derivation yields the norm
\begin{equation}\label{reQTStatenorm-c2}
	\begin{aligned}
\mathcal{N}_{\varphi_R}(t_0)=\langle\psi_R(t_0| \theta^2|\psi_R(t_0)\rangle + 16\pi^2\langle \psi_f(t_0)| \psi_f(t_0)\rangle-8\pi\text{Im}\left[\langle \psi_f(t_0)|\theta|\psi_R(t_0)\rangle  \right]\:,
	\end{aligned}
\end{equation}
where the $\text{Im}[\cdot]$ indicates the imaginary part of a complex variable.

The first term in the right side of Eq.~\eqref{reQTStatenorm-c2} can be approximated as
\begin{equation}\label{reLQuSNfirst-c2}
	\begin{aligned}
\langle\psi_R(t_0| \theta^2|\psi_R(t_0)\rangle=(K^2+\lambda^2) t_n^2\int_{-\pi}^{\pi}\theta^2\sin^2(\theta)\exp{\left[\frac{\lambda t_n}{2\pi}\cos(\theta)\right]}|\psi(\theta,t_n)|^2d\theta
 \approx 3\pi^2\frac{(K^2+\lambda^2)}{\lambda^2}I_0\left(\frac{\lambda t_n}{\pi}\right)\:.
	\end{aligned}
\end{equation}
The second term is
\begin{equation}\label{reQuSNsecond-c2}
	\begin{aligned}
16\pi^2\langle \psi_f(t_0)| \psi_f(t_0)\rangle=16\pi^2\int_{-\pi}^{\pi}\exp{\left[\frac{\lambda t_n}{2\pi}\cos(\theta)\right]}|\psi(\theta,t_n)|^2d\theta=16\pi^2 I_0\left(\frac{\lambda t_n}{\pi}\right)\:.
	\end{aligned}
\end{equation}
And, the third one is
\begin{equation}\label{reQuSNthird-c2}
	\begin{aligned}
8\pi\text{Im}\left[\langle \psi_f(t_0)|\theta|\psi_R(t_0)\rangle  \right]=4\lambda t_n\int_{-\pi}^{\pi}\exp{\left[\frac{\lambda t_n}{\pi}\cos(\theta)\right]}\theta\sin(\theta)d\theta
\approx 8\pi^2I_0\left(\frac{\lambda t_n}{\pi}\right)\:.
	\end{aligned}
\end{equation}
Combing Eqs.~\eqref{reLQuSNfirst-c2},~\eqref{reQuSNsecond-c2}, and ~\eqref{reQuSNthird-c2}, we obtain the expression of the norm
\begin{equation}\label{reQTStnorm-c2}
	\begin{aligned}
\mathcal{N}_{\varphi_R}(t_0)=\pi^2\frac{3K^2+11\lambda^2}{\lambda^2}
I_0\left(\frac{\lambda t_n}{\pi}\right)\:.
	\end{aligned}
\end{equation}

Plugging Eqs.~\eqref{INStatenorm-c2}, ~\eqref{QTStnorm-c2}, and~\eqref{LLLastQuSnormtn-c2} into Eq.~\eqref{numrenormalization-c2t0last} yields the $C_2(t_n)$
\begin{equation}\label{C2LS-c2''}
	\begin{aligned}
C_2(t_n)=\tilde{\mathcal{N}}_{\varphi}(t_n)\frac{\mathcal{N}_{\varphi}(t_0)}{\mathcal{N}_{\varphi}(t_n)}
\approx 2\pi^3\frac{(K^2+\lambda^2)}{\lambda}t_n\:.
	\end{aligned}
\end{equation}

In Hermitian case, i.e., $\lambda=0$, both $\psi_f(\theta,t_0)$ and $\psi_R(\theta,t_0)$ are real. Therefore,
the $C_2(t_n)$ has the expression
\begin{equation}\label{reQTStatenorm-c2-Hermi}
	\begin{aligned}
	C_2(t_n)=\langle\psi_R(t_0| \theta^2|\psi_R(t_0)\rangle + 16\pi^2\langle \psi_f(t_0)| \psi_f(t_0)\rangle=\frac{\pi^2K^2}{6}t_n^2 + 16\pi^2\:.
	\end{aligned}
\end{equation}

\subsection*{S.2.3 Detailed derivation of $C_3(t_n)$}

The $C_3(t_n)$ is defined as
\begin{equation}\label{C3quan-c3-0}
	\begin{aligned}
C_3(t_n)=\langle \psi_R(t_0)|\theta|\varphi_R(t_0)\rangle=\langle \psi_R(t_0)|\theta^2|\psi_R(t_0)\rangle-i 4\pi\langle \psi_R(t_0)|\theta|\psi_f(t_0)\rangle\:.
	\end{aligned}
\end{equation}
By using Eqs.\eqref{reLQuSNfirst-c2} and \eqref{reQuSNthird-c2}, we can derive its real part
\begin{equation}\label{RLC3quan-c3Last}
	\begin{aligned}
\text{Re}[C_3(t_n)]=\langle \psi_R(t_0)|\theta^2|\psi_R(t_0)\rangle- 4\pi\text{Im}\left[\langle \psi_f(t_0)|\theta|\psi_R(t_0)\rangle\right]\approx \pi^2\frac{3K^2-\lambda^2}{\lambda^2}I_0\left(\frac{\lambda t_n}{\pi}\right)\:.
	\end{aligned}
\end{equation}
Applying Eq.\eqref{TPart-Num}, the rescaled $\text{Re}[C_3(t_n)]$ can be written as
\begin{equation}\label{intheoryRLC3quan-c3L}
	\begin{aligned}
\text{Re}[C_3(t_n)]=\text{Re}[C_3(t_n)]\sqrt{\frac{\mathcal{N}_{\psi}(t_0)\tilde{\mathcal{N}}_{\psi}(t_n)}{\mathcal{N}_{\psi}(t_n)\mathcal{N}_{\psi_R}(t_0)}}
\sqrt{\frac{\mathcal{N}_{\varphi}(t_0)\tilde{\mathcal{N}}_{\varphi}(t_n)}{\mathcal{N}_{\varphi}(t_n)\mathcal{N}_{\varphi_R}(t_0)}}=\frac{\pi^2(3K^2-\lambda^2)}{\lambda^2}\sqrt{\frac{4\pi\lambda(K^2+\lambda^2)}{3K^2+11\lambda^2}}\cdot\sqrt{t_n}\:.
\end{aligned}
\end{equation}

In Hermitian case, the $C_3(t_n)$ has the expression
\begin{equation}\label{RLC3quan-c3-Hermi}
	\begin{aligned}
\text{Re}[C_3(t_n)]=\langle \psi_R(t_0)|\theta^2|\psi_R(t_0)\rangle=\frac{K^2}{2\pi}t_n^2\int_{-\pi}^{\pi}\theta^2\sin^2(\theta)d\theta=\frac{\pi^2K^2t_n^2}{6}\:.
	\end{aligned}
\end{equation}

Combining Eqs.~\eqref{C1LS-e2}, ~\eqref{C2LS-c2''}, and~\eqref{intheoryRLC3quan-c3L} yields the expression of the OTOCs for $\lambda t_n\gg 1$
\begin{align}\label{otoc-overal-NH}
\begin{aligned}
C_p(t_n)=C_1(t_n)+C_2(t_n)-2\text{Re}[C_3(t_n)]\approx 2\pi^3\frac{(K^2+\lambda^2)}{\lambda}t_n\:,
\end{aligned}
\end{align}
where we neglect the contributions from both the constant term $C_1$ and the square root of $t_n$ term of $C_3$, as they are significantly smaller than $C_2$ for $t_n\gg 1$.
Based on the relations in Eqs.\eqref{C1Hermitian},~\eqref{reQTStatenorm-c2-Hermi}, and~\eqref{RLC3quan-c3-Hermi} in the Hermitian case ($\lambda=0$), we can conclude that
\begin{align}\label{otoc-overal-H}
C_p(t_n)=16\pi^2\;.
\end{align}

\begin{figure}[b]
\begin{center}
\includegraphics[width=12cm]{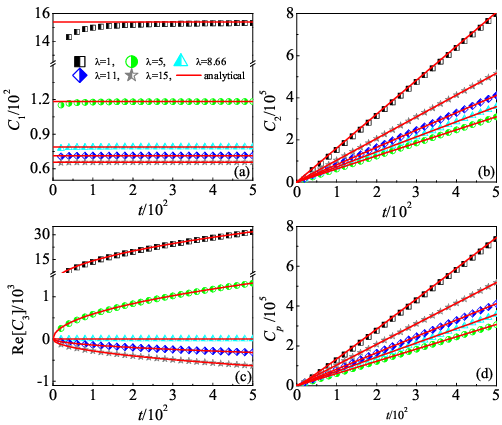}
\caption{Time dependence of $C_1$ (a), $C_2$ (b), $\text{Re}[C_3]$ (c), and $C_p$ (d)  with $\lambda=1$ (squares), $5$ (circles), $8.66$ (triangles), 11 (diamonds), and 15 (pentagrams). The red lines in (a), (b), (c), and (d) indicate our theoretical prediction in Eq.~\eqref{C1LS-e2},~\eqref{C2LS-c2''},~\eqref{intheoryRLC3quan-c3L}, and~\eqref{otoc-overal-NH}. The parameters are $K=5$ and $\ehbar=4\pi$.\label{TPrtOTC}}
\end{center}
\end{figure}

To verify theoretical analysis above, we numerically investigate the time dependence of the $C_1$, $C_2$, $\text{Re}[C_3]$, and $C_p$ for different $\lambda$. Figure~\eqref{TPrtOTC}(a) demonstrates that $C_1$ rapidly increases to saturation over time for a non-zero $\lambda$ (e.g., $\lambda=1$). Furthermore, the saturation value agrees well with Eq.\eqref{C1LS-e2}. In the non-Hermitian case [e.g., $\lambda=1$ in Fig.\eqref{TPrtOTC}(b)], $C_2$ linearly increases with time, following the laws described by Eq.\eqref{C2LS-c2''} perfectly. Interestingly, our numerical results in Fig.\eqref{TPrtOTC}(c) demonstrate that $\text{Re}[C_3]$ increases (or decreases) as the square root of $t$ for $\lambda$ smaller (or larger) than a threshold value of $\lambda_c \approx 8.66$, and is almost zero when $\lambda=\lambda_c$. The threshold value is determined by Eq.\eqref{intheoryRLC3quan-c3L}, specifically $3K^2-\lambda^2=0$, which confirms the validity of our theoretical prediction.  Additionally, the $C_p$ exhibits linear growth with time for non-zero $\lambda$ and is approximately equal to $C_2$, as predicted by Eq.\eqref{otoc-overal-NH} [see Fig.~\eqref{TPrtOTC}(d)].

\end{document}